\documentclass[
]{ceurart}

\usepackage{listings}
\usepackage{makeidx}         
\usepackage{graphicx}        
\usepackage{multicol}        

\usepackage{newtxtext}       %
\usepackage{newtxmath}       

\usepackage{booktabs}
\usepackage[english]{babel}
\usepackage{amssymb}
\usepackage{url}
\usepackage{xcolor}
\usepackage[export]{adjustbox}
\usepackage{colortbl}
\makeatletter
\def\hlinewd#1{%
  \noalign{\ifnum0=`}\fi\hrule \@height #1 \futurelet
   \reserved@a\@xhline}
\makeatother

\begin{document}

\copyrightyear{2025}
\copyrightclause{Copyright for this paper by its authors.
  Use permitted under Creative Commons License Attribution 4.0
  International (CC BY 4.0).}

\conference{Preprint / Artificial Intelligence and Cognition (AIC 2025)}

\title{A Human-Factors Guided Cognitive Model of\\Visuospatial Complexity in Embodied Active Vision}


\author[1,4]{Vasiliki Kondyli}[%
orcid=0000-0003-0392-026X,
email=vasiliki.kondyli@psy.lu.se,
]
\address[1]{Lund University, Sweden}

\author[2,4]{Mehul Bhatt}[%
orcid=0000-0002-6290-5492,
email=mehul.bhatt@oru.se
]
\address[2]{Örebro University, Sweden}

\author[3,4]{Jakob Suchan}[%
orcid=0000-0002-1356-3297,
email=jsuchan@constructor.university,
]
\address[3]{Constructor University Bremen, Germany}

\address[4]{CoDesign Lab EU., \href{https://codesign-lab.org/cognitive-vision/}{codesign-lab.org/cognitive-vision}}

\begin{abstract}
We propose a novel framework for the analysis of multimodal data --encompassing visual, auditory, and spatial stimuli-- foregrounding the role of complexity in embodied perception and interaction in dynamic, naturalistic settings. Grounded in theories of embodied cognition and active vision, we argue that embodied perceptual complexity emerges from an agent's dynamic engagement with the environment and must be analyzed holistically, as a combination of qualitative and quantitative attributes pertaining to, for instance, visuospatial and auditory features. Building on previous work on visual complexity, we expand this into a categorization of diverse complexity attributes—quantitative, structural, dynamic, auditory, and interactional—that together characterize multimodal complexity. We demonstrate how this model provides a theoretical framework for characterizing aspects of visuospatial complexity and their interactions, specifically in the context of everyday driving. We also discuss practical applications of the proposed model for creating and evaluating  benchmark datasets (e.g., in driving) that centralize cognitive human factors, as well as applications aimed at systematically investigating the effect of visuospatial complexity on human active vision from the viewpoint of visual perception research. The proposed framework lays the foundation for automated methods that interpret complexity in 3D dynamic environments from a human-centered perspective, serving as a semantic template for explainable computational analysis of visuospatial complexity with a categorical focus on cognitive human factors.
\end{abstract}


\maketitle

\section{Introduction}

As assistive technologies and AI systems increasingly integrate multimodal data—such as visual, auditory, spatial inputs—traditional models of visual complexity, typically based on static and unimodal data, are no longer sufficient. This paper introduces a new framework for understanding complexity in multimodal contexts, emphasizing the role of embodied perception and interaction in dynamic, naturalistic environments. Grounded in theories of embodied cognition and active vision, we argue that visuospatial complexity arises from an agent’s active engagement with the environment and must be analyzed holistically as a combination of qualitative and quantitative attributes. Building on previous work, we categorize the attributes into several groups: quantitative, structural, dynamic, auditory, and interactive. We demonstrate how this model provides a theoretical framework for characterizing visuospatial complexity in naturalistic circumstances, focusing on everyday driving as a context with technological interest. In addition, we discuss practical applications of the model, including its use in designing and evaluating driving benchmark datasets with a focus on human factors, as well as in systematically studying the effects of visuospatial complexity on human active vision through interactive virtual reality experiments. Ultimately, this framework lays the groundwork for developing automated methods to interpret complexity in dynamic 3D environments from a human-centered perspective, serving as a semantic foundation for explainable computational analysis of visuospatial complexity.

\medskip

\textbf{The Human-Centred Imperative in Complex Dynamic Environments.} \quad  Emerging AI and assistive technologies increasingly operate in dynamic, multimodal, and socially structured environments. Whether it is autonomous vehicles navigating densely populated urban spaces, augmented or virtual reality systems overlaying information in complex real-world scenes, or social robots engaging in natural interaction with humans, these systems must move beyond static scene analysis. In other words, these systems must function in dynamic, real-world environments where perception is situated, interactive, and dependent on a continuous loop between action and sensory feedback. To do so they need to account for how visual, auditory, spatial, and temporal cues interact to influence human perception and action. It is no longer sufficient for computational tools to recognize what is in a scene—they must understand how the environment is experienced and navigated by embodied agents. This requires capturing the temporal unfolding of sensory experiences and the fluid interplay between action and perception. Drawing on embodied cognition research, which emphasizes the inseparability of perception and action in real-world behavior \cite{Dourish2001,Clark2013}, we argue that modeling environmental complexity demands a shift toward representations that reflect human-centered, interactive, and embodied experiences in real-time. This paper argues for the development of a richer framework for characterizing environmental complexity, one that captures how multimodal cues—visual, auditory, spatial, and temporal—shape human behavior and cognition during active, embodied tasks. Central to this shift is the concept of active vision:

\begin{quote}
{\small\sffamily
\noindent
\textbf{Active Vision} \quad in cognitive psychology, describes a purposeful, embodied process in which the agent explores the environment through movement and interaction \cite{Findlay2003,Tatler2011}.}
\end{quote}

As AI agents increasingly engage in real-world tasks, active vision becomes critical for interpreting the social and spatial nuances of their surroundings. Now, more than ever, the need to model how complexity arises from the interplay of multimodal signals and embodied activity is urgent: without this, AI systems will struggle to meet human-centered expectations and operate responsibly in complex, everyday environments. This rethinking of environmental complexity is foundational for enabling AI systems to understand not just what is present in a scene, but how it matters for behavior, intention, and interaction.

\medskip

\textbf{Human-Centred Benchmarking and Standardisation.} \quad Ensuring that future autonomous systems—such as self‑driving vehicles—can operate safely and intuitively around humans requires more than mastering steering and speed control it demands a deeper understanding of human‑environment interaction. While large multimodal datasets like nuScenes and Waymo Open Dataset  \cite{nuScenes2021,Waymo2021} offer extensive sensor coverage, they lack the human‑centred annotations that capture how multimodal cues and embodied interactions shape scene complexity. At the same time, regulatory frameworks—such as Germany’s BMVI  \cite{bmvi2020} set of 20 key propositions—emphasize ethics, transparency, and human rights, yet offer little guidance on technical standards grounded in human perception, attention, and interaction in dynamic environments. By enriching these datasets with our visuospatial complexity model, we provide a foundation for evaluating whether autonomous systems encounter and learn from the full range of real‑world complexity—spanning visual clutter, motion dynamics, interaction density, and audio-visual context. This integrated benchmark enables complexity‑aware training regimes, systematic data coverage analysis, and reproducible empirical testing. It also supports higher‑level cognitive tasks—such as scene understanding, commonsense reasoning, and multimodal question answering—by linking low‑level signals with semantic, interpretive annotations. Ultimately, this human‑centred approach elevates standardisation in autonomous systems by aligning technical performance with ethical, perceptual, and social expectations in naturalistic settings.

\medskip

\textbf{Key Contributions.} \quad We present an extended cognitive model of visuospatial complexity tailored to dynamic, real-world driving scenes. Building on prior work, our model integrates both low-level visual attributes (e.g., object bounding boxes) and higher-level semantic annotations to more fully capture the richness of naturalistic environments. We systematically organize scene characteristics into five interrelated categories: quantitative, structural, dynamic, auditory, and interactive. Our findings demonstrate that visuospatial complexity is not solely driven by visual clutter; rather, it emerges from interactions among multiple attributes. Notably, high complexity can arise from combinations of features that, in isolation, may be only moderately demanding. As a proof of concept, we apply the model to construct a benchmark dataset of annotated driving scenes, illustrating how the interplay of these dimensions contributes to perceived complexity. We also use the model to generate diverse virtual environments corresponding to varying levels of visuospatial complexity, which we test in an embodied, multimodal virtual reality (VR) study of active human vision. Results show that complexity—defined according to our model—significantly influences both attentional dynamics and driving performance. This work lays the foundation for cognitively grounded benchmarks and tools that integrate human factors, enabling human-centered evaluation and interpretable visual analysis in autonomous driving systems.

\section{Modelling Visual Complexity}

\begin{figure}[t]
	\centering
        \includegraphics[width=\textwidth]{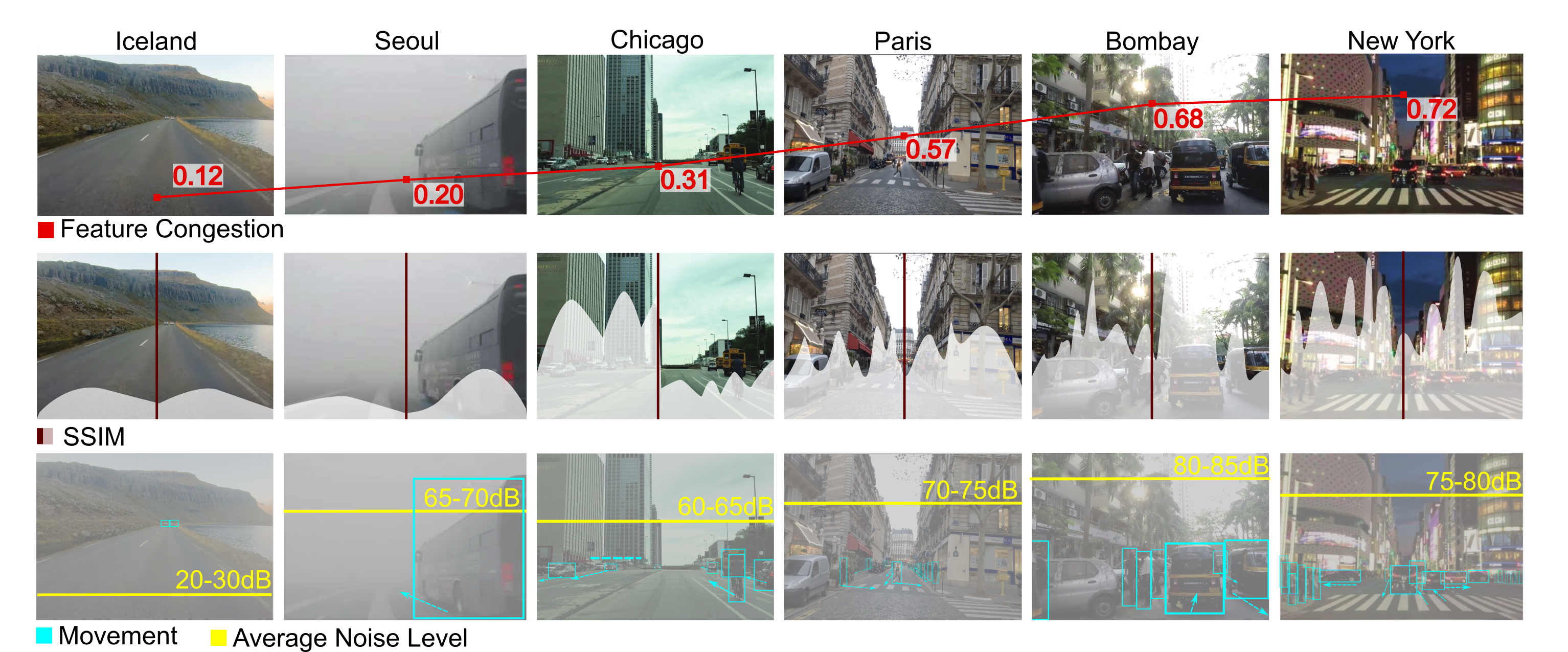}
	\caption{Examples of commonly used visual complexity metrics from the literature, illustrated through their application to sample driving scenes.} 
\label{fig:scenes}
\end{figure}

Much of the foundational research on scene complexity stems from visual perception studies and has been shaped by diverse disciplines including cognitive science, psychology, computer science, and marketing \citep{Cavalcante2014,Moacdieh2015}. \textbf{Visual Complexity} is typically defined as the degree of detail, variability, or intricacy within a scene \citep{Snodgrass1980} and is commonly measured using both computational and behavioral methods that assess features such as clutter, object density, and scene entropy \citep{Rosenholtz2007,Henderson2009,Madan2018}. However, these approaches often focus on static, two-dimensional imagery and overlook the embodied, dynamic, and multisensory nature of perception in real-world environments. To better reflect the perceptual demands of everyday cognition—such as active vision, spatial navigation, and interactive behavior—we propose the broader construct of \textbf{Visuospatial Complexity}. This term captures the integration of visual and spatial characteristics as they co-exist in dynamic, naturalistic scenes where people  act. In what follows, we review the state of the art in traditional and recent measures of scene complexity, as well as recent research from human factors and applied cognition that, while not always framed explicitly in terms of visuospatial complexity metrics, provides valuable insight into how people perceive, interpret, and act within complex environments (Fig. \ref{fig:scenes}). Our aim is to build an inclusive framework that accounts for the temporally unfolding, three-dimensional nature of visuospatial experience, where perception and action are tightly coupled.

\textbf{Low-Level Scene Features.} \quad This category captures how visually dense and perceptually demanding a scene is, focusing on the early stages of visual processing. Measures such as feature congestion, which combine variability in color, contrast, and orientation, have been shown to reliably predict perceived clutter and search difficulty across natural scenes \cite{Rosenholtz2005,Rosenholtz2007}. Complementary metrics like edge density, texture entropy, and subband entropy quantify spatial irregularity and encoding complexity, especially in wavelet-transformed images \cite{BravoFarid2008, Semizer2019}. These low-level computational features are grounded in psychophysical evidence that links visual clutter with increased cognitive load and reduced visual search efficiency, particularly in applied contexts such as driving or interface use \cite{Neider2008,Smith2013}. The notion of clutter broadly encompasses the number, variety, and density of visual elements from the viewer’s perspective, all of which contribute to perceptual load. These features not only influence attentional allocation but also constrain object recognition and scene interpretation. Methods like feature congestion and subband entropy have been extended to incorporate semantic information and validated using human behavioral and eye-tracking data \cite{Henderson2009,Madan2018}. Collectively, these models reflect the fundamental link between early visual attributes and perceptual performance, offering reliable predictors for how spatial complexity affects attention and behavior.

\textbf{Structural Organization.} \quad Structural attributes describe how visual elements are arranged in space—through alignment, grouping, and overall spatial distribution and reflect how well-organized or perceptually coherent a scene is. Regular spatial configurations, such as symmetry or orderly layouts, reduce perceived complexity and support faster scene interpretation  \cite{Gartus2017}. Metrics such as the Structural Similarity Index (SSIM) quantify this organization by comparing luminance, contrast, and structural patterns between images \cite{Wang2004}. Its extension, MS-SSIM, incorporates multiple spatial scales to reflect both fine and coarse structural coherence  \cite{Wang2003}. These indices, though initially designed for image quality assessment, are now applied to perception studies to assess how structure influences attention and recognition.

\textbf{Scene Dynamics and Semantics.}  \quad When assessing the complexity of dynamic, real-world stimuli—such as videos, interactive environments, or naturalistic scenes—it is essential to move beyond static visual features and incorporate multiple interacting dimensions. Temporal dynamics such as motion, flicker, and observer movement increase perceptual load and attentional demands. Computational metrics like Temporal Information (TI) and optical flow capture these dynamic properties and help quantify visual change over time \citep{Mital2011, Warren2004}. Auditory cues also contribute to perceived complexity by providing additional layers of information that influence arousal, spatial orientation, and attentional shifts. Metrics such as spectral entropy, the Acoustic Diversity Index (ADI), and the Acoustic Evenness Index (AEI) measure variability and richness in environmental soundscapes \citep{Sueur2014, Villanueva-Rivera2011, Weisser2019}, offering insight into how complex auditory scenes shape perception. Crucially, dynamic real-world stimuli are characterized by rich high-level semantic structure. Analyzing such interactions requires systematic capture across multiple modalities—including visual, spatial, and auditory channels—along fine-grained annotations of events, body movements, and gaze behavior within contextually rich scenes \citep{Nair-ACPTAP-2026,Nair-dataset}. These multimodal tools enable the investigation of how individuals attend to socially meaningful events, such as joint attention and turn-taking, and how such events influence cognitive processes including event segmentation, interpretation, and memory formation.


 \subsection{Advancing the Cognitive Model of Visuospatial Complexity}


 Most existing models and metrics capture isolated aspects of complexity—such as visual clutter, structural coherence, dynamic motion, or auditory cues. While some newer approaches incorporate audio, temporal dynamics, or social and semantic context, they often remain limited to static 2D scenes and object-level analysis. Critically, they overlook the multimodal and embodied nature of real-world perception. Real environments are dynamic, multisensory, and socially interactive—experienced through active vision and physical engagement. Addressing perceptual dimensions in isolation leads to a fragmented understanding of human experience. To overcome this, we propose a model of visuospatial complexity that takes a holistic, multimodal approach, integrating insights from spatial cognition, perception, and interactive behavior.  This broader perspective allows for a more systematic understanding of how environmental attributes shape behavior and enables the development of intelligent systems capable of both evaluating human experience and predicting future responses.

 \begin{table}
\centering
\scriptsize
\arrayrulecolor[gray]{0.8}


\begin{tabular}{
    >{\columncolor[gray]{0.92}}p{3.3cm}|p{1.4cm}|p{1.2cm}|p{1.2cm}|p{0.9cm}|p{1.4cm}|p{4cm}|}
\hlinewd{1pt}
\rowcolor[gray]{0.92}\textbf{Metric / Model} & \textbf{Quantitative} & \textbf{Structural} & \textbf{Dynamic} & \textbf{Audio} & \textbf{Interaction} & \textbf{Details} \\
\hline
Feature Congestion \cite{Rosenholtz2007}  &  \centering X &  &  &  &  &  edge density, color contrast, and orientation variability \\
\hline
Subband Entropy \cite{Rosenholtz2005}  &  \centering X &  &  &  &  & spatial frequency and texture variability via wavelet transforms\\
\hline
Perceptual Clutter Model \cite{BravoFarid2008} &  \centering X &  \centering X &  &  &  & boundaries, textures, segmentation  \\
\hline
Structural Similarity (SSIM) \cite{Wang2004}  &  &  \centering X &  &  &  & patch structure, multi-scale and temporal similarity   \\
\hline
 Two-component computational model \cite{Gartus2017} &  \centering X &  \centering X &  &  &  &  element count, mirror symmetry \\
 \hline
Optic flow, Flicker  \cite{Mital2011,Warren2004}  &  \centering X  &  &  \centering X &  &  & number of elements, luminance, direction, speed of movement \\
 \hline
Acoustic Diversity \& Evenness Index \cite{Villanueva-Rivera2011} &  &  &  & \centering X &  & richness, balance \\
 \hline
Multimodal Interactions  \cite{Nair-ACPTAP-2026,Nair-dataset} &  &  &  \centering X &  \centering X & \centering X &  scene semantics, gestures, attention temporal dynamics\\

\hline

\hline
\end{tabular}

\caption{State-of-the-art metrics addressing different aspects of scene complexity.}
\label{tab:complexity_models}
\end{table}


\section{Human-Centred Complexity Model for Active Vision}

Drawing on principles from embodied cognition and active vision, our framework emphasizes the dynamic, egocentric, and goal-driven nature of perception. Rather than treating complexity as a static attribute of the environment, we view it as emergent—arising from the continuous interaction between the observer and their context. Prior models have largely emphasized static visual features such as edge density, object count, or feature congestion, often neglecting the temporal, auditory, and social dimensions that characterize real-world dynamic experience. Our model addresses these gaps by integrating spatial, temporal, and multimodal cues to capture complexity as it unfolds over time.

\subsection{Defining a Visuospatial Complexity Model} 

Building on our previous work on defining visuospatial complexity for driving, and navigation  \cite{Kondyli2020STAIRS,KondyliICORD21}, we expand the model to include attributes crucial attributes for cognitive load in streetscapes.   For example, a driving scene may appear visually simple when viewed as a single frame, yet becomes cognitively demanding when experienced as a dynamic sequence involving moving pedestrians, cyclists, head movements, and auditory stimuli. From this perspective, complexity is relational—it is not solely an environmental property, but is co-constructed through the observer’s perceptual capacities, intentions, and interactions. This relational view is particularly important in interactive settings such as driving, where visuospatial complexity evolves in concert with agent behavior and multimodal information flow. To support a more comprehensive account of visuospatial complexity, our framework incorporates both quantitative metrics (e.g., density, flicker, structural similarity) and qualitative attributes (e.g., social coordination, contextual relevance). This dual-layered approach enables a richer characterization of perceptual demands across dynamic, embodied tasks. Ultimately, the framework offers a principled basis for modeling complexity in naturalistic environments and supports the development of shared vocabulary to link low-level sensory inputs with high-level semantic understanding. Based on this approach, we propose an updated taxonomy of visuospatial complexity (\textbf{A1-A5}) as follows (Table \ref{tbl:visual complexity model}):

{
\begin{table}[t]
\renewcommand{\arraystretch}{1.1}
\begin{center}
 \scriptsize\sffamily
\begin{tabular}{>{\columncolor[gray]{0.92}}l p{12.5cm}}
\hlinewd{1pt}
\rowcolor[gray]{.92}\textbf{VISUOSPATIAL COMPLEXITY} & \textbf{Description} \\\hline\hline

\textbf{A1. Quantitative attributes} &  \\
\hline

 Quantity   &  No. components  (objects, people, shapes etc.) \\
 Variety of Colours & No. colours \\
 Variety of Shapes/Objects & No.  shapes / objects   \\ 
 Objects Density &  No. objects in a defined area \\
 Edges Density &  No. edges of objects in a scene / visual area  \\
 Luminance &  Amount of light emitted / reflected from the scene \\
Saliency &  Particularly prominent objects based on characteristics of colour, luminance and contrast   \\ 
Target-background similarity & Compare similarity in luminance, contrast, structure, or spatial and orientation information \\
Size (length-width-height)&  The dimensions of the physical space, the area coved by the visual stimulus.  \\[6pt]

\hline
\textbf{A2. Structural  Attributes}     &  \\
\hline

Repetition & Recurrence of the same element of group of elements or characteristics on a line, a grid  or a patterns in space \\
Symmetry & Resilience to transformation and movement. Types: reflectional, rotational, translational, helical, fractal   \\ 
Order & Organised elements based on a recognised structure, Varies from poorly organised to highly organised \\
Homogeneity/Heterogeneity  & The state of being all the same kind/ diverse. Varies from single shape repeated to multiple distinct shapes  \\  
Regularity  & Variations in a placement rule across a surface or line; Varies from simple polygons to abstract shapes  \\
Openness & The ratio between empty and full space  \\ 
Grouping & No. elements that are part of a group  \\
Rotation Metric  &  Accumulated degrees of rotation angle during locomotion, No. of turns \\
Visibility &  Visual range from a vantage point, visual connectivity between points\\
Interconnection Density & No. of directional choices in each node (e.g. decision point, junction)\\

\hline
\textbf{A3. Dynamic  Attributes}  &   \\
\hline

Motion & No. people or objects moving in the scene \\ 
Flicker   &  Abrupt changes over-time (in luminance, colours, etc.)  \\
Speed | Direction & The rate of change of position with respect to time | Move or facing towards \\

\hline
\textbf{A4. Auditory  Attributes}  &   \\
\hline

Alerting & \quad High-intensity or abrupt sounds that draw immediate attention  (alarms, tire screeches) \\
(In)Congruent & \quad Sound that aligns with the visible context or not (footsteps, turning signal before a car turns, horn)\\ 
Noise & \quad Ambient/background sounds without a clear semantics (urban hum, construction noise, static) \\
Mechanical  & \quad vehicles or infrastructure (engine revving, brakes, bicycle bells) \\
Social  & \quad Verbal or non-verbal interactions between people (greetings, arguments, laughter) \\
Environmental & \quad Characteristics of a setting but not  related to traffic  (birds chirping, water, wind) \\

\hline
\textbf{A5. Interaction  Attributes}  &   \\
\hline

Mode  & \quad   Explicit, or Implicit interactions (hand wave, slowing down) \\
Method   & \quad Medium or mechanism of interaction  (Formal, Informal, Device-based, Body-based)  \\

Social Attention &   \quad  Attentional engagement (Individual, Joint, Monitoring, Mutual)  \\

Modalities & \quad  Sensory channels used to signal intent or react to others (speech, head movements, gestures, etc.) \\
Social Role & \quad Role of the interacting party, as it shapes both expected behavior and vulnerability (pedestian, cyclist, etc.)\\

\hlinewd{1pt}
\end{tabular}
\caption{{\sffamily\footnotesize Taxonomy of attributes for Visuospatial Complexity}}
\label{tbl:visual complexity model}
\end{center}
\end{table}%
}

\medskip

\textbf{A1. Quantitative Attributes.} \quad The physical space and its functional \emph{clutter} are general properties of all scenes that are immediately accessible to humans \cite{Park2014}. In this category we also add the  size of the physical space as it is defined by boundaries in three dimensions:  length, width and height. Recent work from neuroscience suggests that retrosplenial cortex activity patterns are predominantly sensitive to the size of a space, meaning that just a glance at a new space yields sufficient immediate information about its extent  \cite{Park2014}. The size is one of the spatial attributes that is directly encoded and represented and consequently it can constrain our action or navigation in the environment \cite{Hermer2001,Learmonth2002}. 
 
\medskip

\textbf{A2. Structural Attributes.}   \quad  While extensive presence of quantitative attributes can lead to overabundance of information, organising the relation between them in space helps in avoiding information overload. Structural coherence affects perceptual grouping and scene understanding, influencing how efficiently observers can segment and interpret space. For example, high regularity, repetition, symmetry of elements are associated with lower visuospatial complexity \citep{Feldman2004,vanderHelm2000}. Conversely, heterogeneity, regularity of shapes, and openness, also influence complexity \cite{vanderHelm2000,Salingaros2014}.  In the context of embodied locomotion, structural attributes are extended to include locomotive complexity attributes such as the rotation metric (rotational angle), visibility (visual range), and interconnection density (directional choices), which significantly impact the holistic legibility of an environment  \citep{ONeill1991,Underwood2006,Kondyli2018Rotation}.

\medskip

\textbf{A3. Dynamic Attributes.}  \quad To capture the temporal changes within an environment we need to analyse the changes in quantitative attributes of an environment along time, that reflect the rate of complexity as we move in space.  Dynamic attributes have a major impact on visual attention patterns, as clusters of attention often coincide with semantically rich objects such as eyes, hands, etc. \citep{Mital2011}.  So in addition to the metrics of optical flow, and flicker, we also add speed and direction of movement, and temporal variation of the different semantic categories.  These are characteristics of moving objects in the scene and have different dependences on the way they are encoded during a cognitive task  \cite{Carrasco2011,Mital2011}. For instance, motion of a target towards a different direction than the rest objects in the scene promote quicker detection and recognition during visual search \cite{Wolfe2010}.

\medskip

\textbf{A4. Auditory Attributes.}  \quad We analyze semantically relevant sounds, as  significantly shape perception, attention, and cognitive load in complex environments. Congruent sounds enhance reaction times and situational awareness, while incongruent or unexpected sounds can trigger surprise or disrupt task performance \cite{Grenzebach2022,Parmentier2014} In driving, relevant auditory signals—like sirens or warning tones—improve response accuracy, but their effectiveness declines under high cognitive load \cite{Jin2024}. Integrating metrics of perceived loudness, eventfulness, together with semantic relevance of the auditory cue is essential for capturing the multimodal nature of real-world, as for example in streetscape \cite{Verma:2020aa}.

\medskip

\textbf{A5. Multimodal Interaction Attributes.}  \quad encompass a wide range of verbal and non-verbal communication modalities that people employ to provide and interpret intentions in complex dynamic environments. These interactions are highly varied and can convey different meanings depending on the task, environment, or social and dynamic factors. Examples include various forms of communication like gestures (e.g., emblematic, iconic, deictic, beat), head movements (e.g., turning towards the street, tilting, nodding), facial expressions (e.g., smiles, frowns, eye rolling), body postures (e.g., crossing arms, leaning towards a car), and gaze (e.g., eye contact, seeking attention, following another's gaze) \cite{Kondyli2020DHM}.

\subsection{Multimodal Scene Analysis with Visuospatial Complexity Profiling} \label{sec:profiling}


The visuospatial complexity model is designed not only as a theoretical framework for characterizing the complexity of dynamic, naturalistic environments but also as a practical tool to support computational scene analysis. It contributes to ongoing developments in computer vision and AI, particularly those focused on perceptual sensemaking—where commonsense, spatial, and temporal reasoning must be grounded in multimodal inputs \cite{Suchan2021}. To operationalize the model, we derive a structured taxonomy of attributes spanning five core dimensions: quantitative, structural, dynamic, auditory, and interactional. This taxonomy provides a unified language for systematically annotating complex scenes, guiding low-level and high-level annotations most relevant to human perception and cognition. As a example case, we apply this framework to the analysis of naturalistic driving videos, emphasizing the embodied, multimodal experience of the driver, who engages in active vision and continuous interaction with the environment. By capturing both low-level perceptual cues and high-level semantic structures as they evolve over time, the model supports a human-centered analysis of scene dynamics. This approach facilitates the interpretation of environmental changes and their cognitive implications, bridging theoretical insights with practical applications in cognitive technologies and human-centered AI.

In practice, we systematically analyse the videos both quantitatively (via image-based metrics like clutter and SSIM) and qualitatively (through bounding-box annotations, dynamic interaction events). Human evaluators conducted qualitative analysis involving two forms of annotations:
\medskip

\textbf{i. Low-level visuospatial metrics.} Bounding box annotations of visual elements (e.g., vehicles, pedestrians, traffic signs) in HumanSignal \footnote{HumanSignal - graphical image annotation tool (https://humansignal.com/)} to calculate metrics such as edges density, color variation, and structural similarity.  
\medskip

\begin{figure}
	\centering
    \includegraphics[width=1\textwidth]{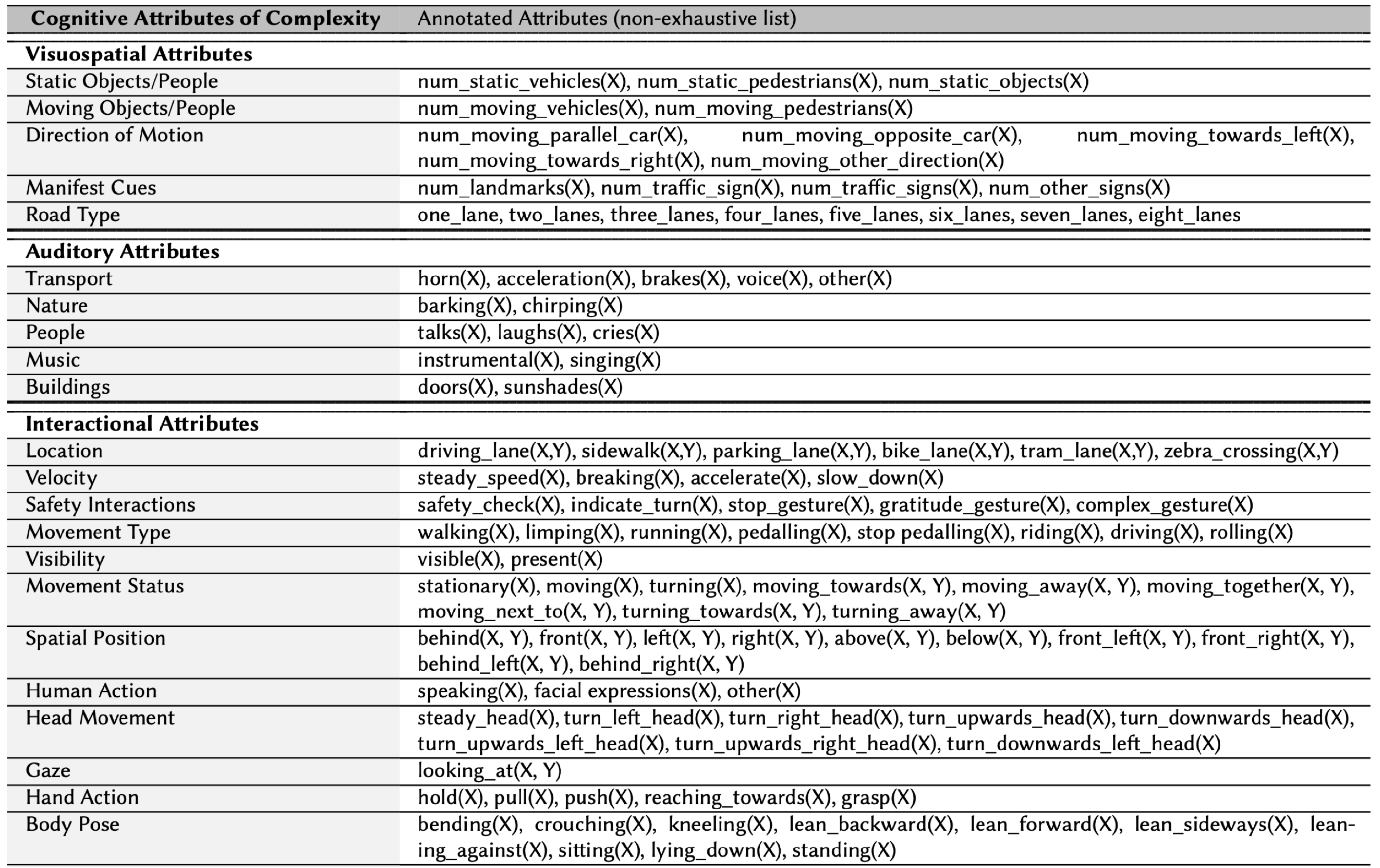} 
	\caption{The taxonomy of visuospatial complexity attributes used for the analysis of real-world dynamic stimuli of the driving dataset.}
	\label{fig:annotations}
\end{figure}

\textbf{ii. High-level visuospatial metrics.} Qualitative scene annotations in ELAN \footnote{ELAN - annotation tool for audio and video recordings (https://archive.mpi.nl/tla/elan)}, capturing multimodal cues (e.g., motion, auditory signals, interaction types) to interpret higher-level semantic content based on the controlled vocabulary of  Fig. \ref{fig:annotations}, pertaining to visuospatial complexity attributes. The result of this qualitative analysis, for the sample scenes in Fig. \ref{fig:driving}, can be summarized  as follows:

  \begin{quote}
{\small\sffamily
\noindent
\textbf{Paris scene:} \quad A Cyclist is {\color{blue!80!black}located on} the right lane in two-lanes road, while Pedestrian A is {\color{blue!80!black}located on} the left lane in front of the Driver, {\color{blue!80!black}crossing} from left to right. Then the Cyclist {\color{blue!80!black}moves to} the left lanes in front of the Driver, and the right lane is  {\color{blue!80!black}occupied} by parked Cars, while the Bus is party  {\color{blue!80!black}located on} the left lane, and Pedestrian B is {\color{blue!80!black}crossing} the lane of opposite traffic from right to left. The Cyclist  {\color{blue!80!black}slows down} as the Bus is  {\color{blue!80!black}approaching}, and Pedestrian 1 is crossing the street from left to right. Then the Cyclist  {\color{blue!80!black}moves towards} the right lanes while a Motorcyclist  {\color{blue!80!black}honks} the horn and then  {\color{blue!80!black}overtakes} the Driver from the left side.}
\end{quote}

  \begin{quote}
{\small\sffamily
\noindent
\textbf{Tokyo scene:} \quad A Driver is {\color{blue!80!black}located on} the right lane in two-lanes road, while many Cars are {\color{blue!80!black}located on} in front of the Driver as well as on the left lane. The Driver is {\color{blue!80!black}approaching} a zebra crossing, while Car A is crossing the zebra crossing and a Van on the left lane {\color{blue!80!black}stops} in front of the zebra crossing. When the Driver {\color{blue!80!black}approaches} the zebra crossing, Car is {\color{blue!80!black}located on} the zebra crossing with the direction left, while many Pedestrians are {\color{blue!80!black}visible} on the side of the Van, {\color{blue!80!black}located on} the left sidewalk in front of the zebra crossing. Then Driver {\color{blue!80!black}stops} in front of the zebra crossing and the Pedestrians {\color{blue!80!black}cross} from left to right the zebra crossing.}
\end{quote}

\textbf{A comparative analysis.} \quad Based on the summarization of the visuospatial complexity analysis of the two scenes, we can compare them based not only on scene variety, but on the richness and interplay of human-relevant perceptual factors. To illustrate this, we present a comparative analysis of two scenes in our benchmark:

\begin{itemize}
\item The \textbf{Tokyo scene} shows high visual clutter and low SSIM, indicating structural density and image variability, but contains relatively few interaction events. 
\item The \textbf{Paris scene} has low clutter and higher SSIM, indicating well-organized and  low-crowed scene, but involves dense sequences of unpredictable pedestrian behavior, rapid motion changes, and high levels of task-related audio.
\end{itemize}

While traditional quantitative metrics might characterize the Tokyo scene as more complex due to visual density, our model reveals that the temporal and interactional richness of the Paris scene increases the level of visuospatial complexity that can lead to a greater cognitive demand. This example highlights that visuospatial complexity in dynamic scenes where people actively interact with their surrounding is not merely a visual property, but an emergent, multimodal phenomenon shaped by embodied perception and the dynamics of human-human and human-environment interaction. Moreover, this finding underscores a key insight, that complexity is context-sensitive and multidimensional. Traditional models, which emphasize surface-level visual features, miss the deeper structure of perceptual challenges that real-world agents must navigate. The suggested holistic complexity profiling—accounting for quantitative as well as qualitative and social dynamics—offers a more ecologically valid approach to evaluating driving scenes.

\begin{figure}[t]
\centering
\includegraphics[width=1\textwidth]{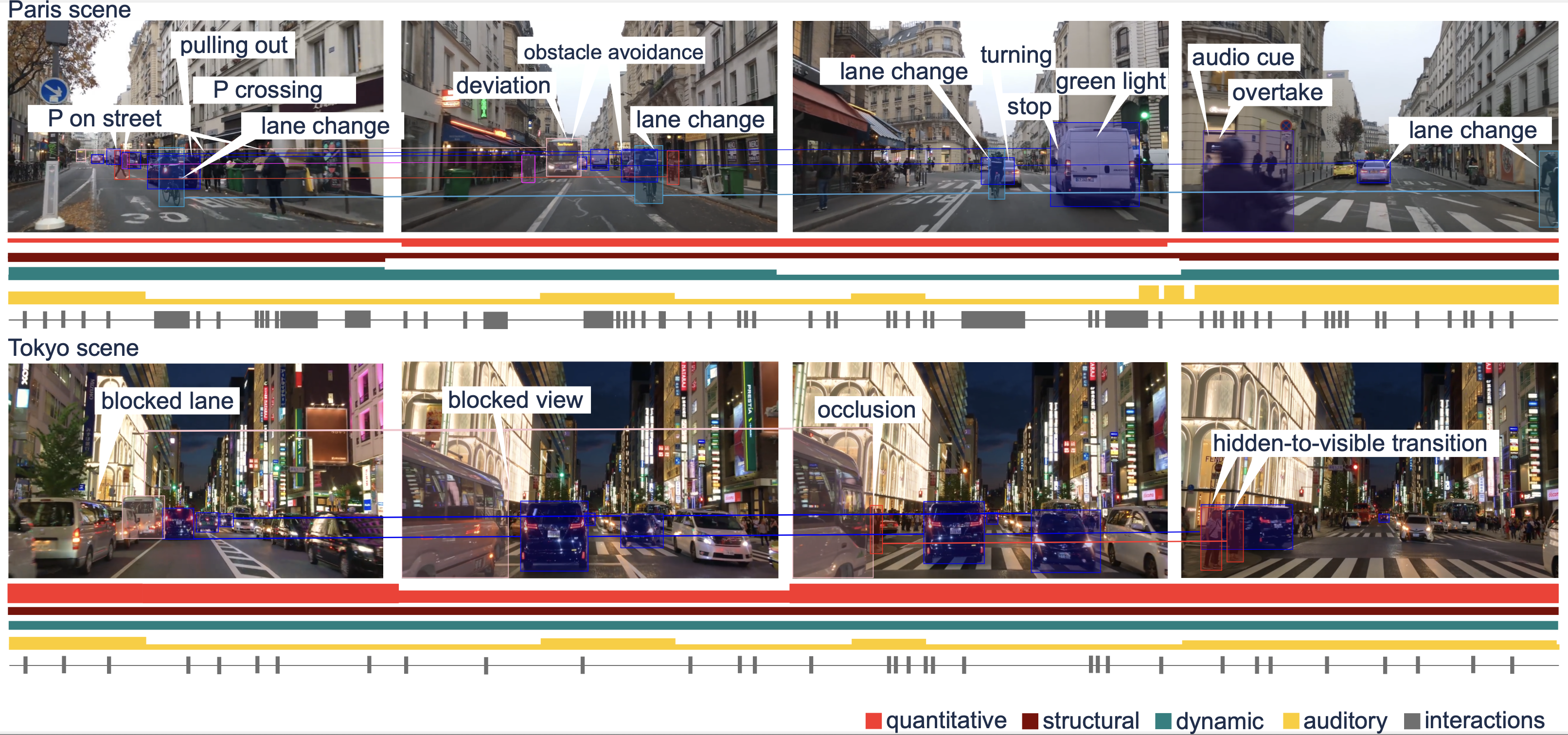}
\caption{Comparing two driving scenes from Paris and  Tokyo, and their (summarized) annotations in respect to quantitative, structural, dynamic, auditory and interactional attributes during 1 min interval.}
\label{fig:driving}
\end{figure}


%

\section{Application the Complexity Model: Dataset and System Evaluation}\label{sec:applications}

The visuospatial complexity model supports both the design and evaluation of cognitive and perceptual technologies in dynamic, naturalistic environments. We demonstrate its utility in two complementary applications. First, we use the model to guide the creation of a benchmark dataset for complexity-based scene analysis in human-centered autonomous driving. Second, we apply the model to the design of a VR driving study, enabling systematic testing of behavior and system performance across environments with varying levels of complexity. In both cases, grounding the analysis in human factors ensures that perceptual and cognitive demands are meaningfully represented—providing a principled foundation for the development and assessment of intelligent systems.

\subsection{Benchmark Dataset Creation and Complexity-Based Scaling}

The visuospatial complexity model provides a structured framework for evaluating AI systems across a broad spectrum of real-world environments, capturing not only visual density but also interactional, multimodal, and cognitive demands. This perspective aligns with recent advancements in perceptual sensemaking in AI, which emphasize the need for commonsense, spatial, and temporal reasoning grounded in multimodal input \cite{Suchan2021}. To demonstrate the practical utility of our model, we curated a benchmark dataset comprising 27 driving videos sampled from diverse urban and rural environments worldwide (e.g., USA, South Korea, Hong Kong, Iceland, India). These scenes represent a wide array of real-world driving conditions—including variations in weather, lighting, culture, and road infrastructure—to ensure ecological validity. They also encompass complex social and traffic interactions, including safety-critical events involving pedestrians, cyclists, wheelchair users, and other vulnerable road users.

\begin{figure}[t]
\centering
\includegraphics[width=\textwidth]{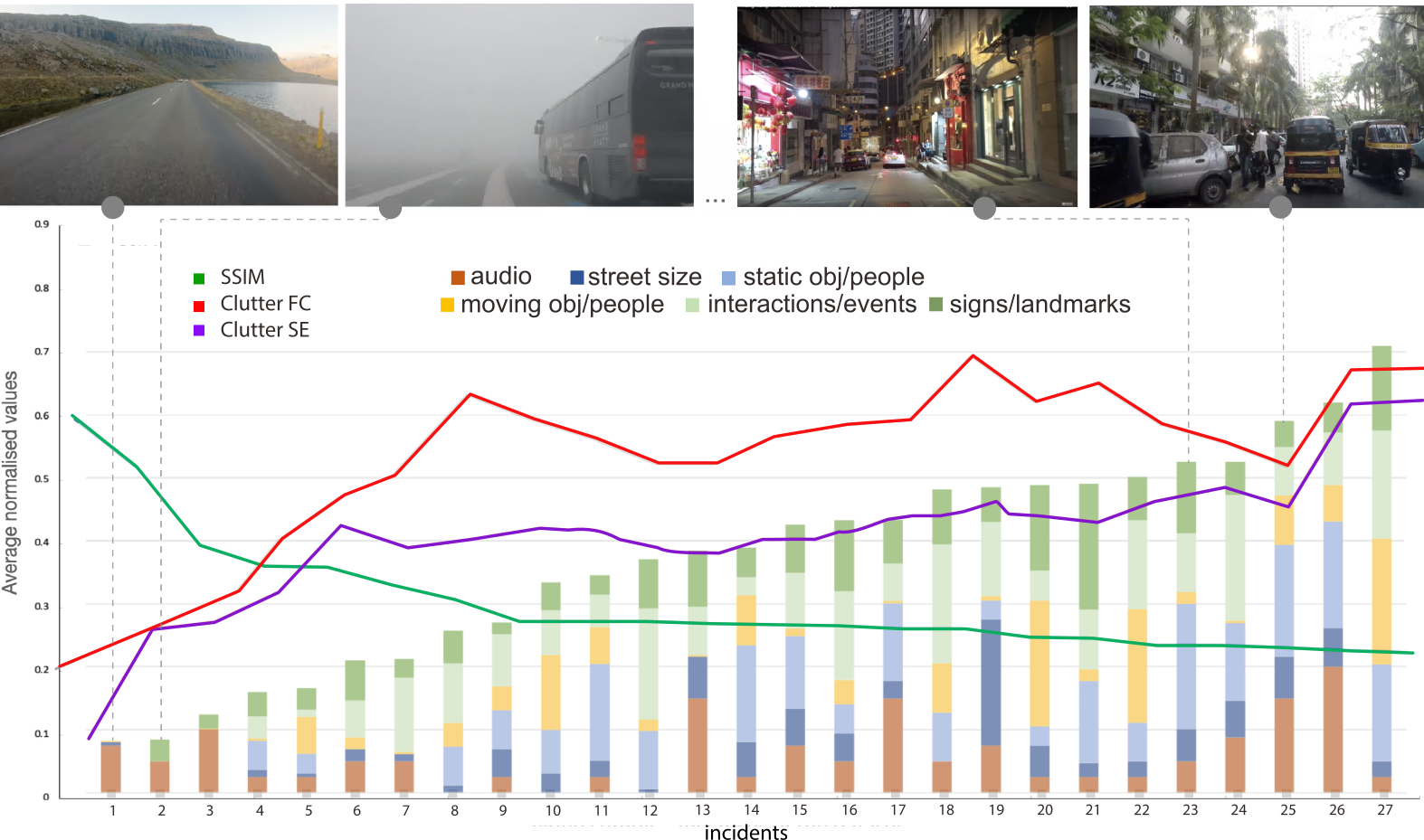}
\label{fig:model_scale}
\caption{The outcome of the analysis of 27 real-world dynamic driving scenes. Each scene is positioned along a graded visuospatial complexity scale, ranging from lower (left) to higher (right) complexity, based on a composite of low-level and high-level attributes defined by the visuospatial complexity model.}
\end{figure}

Each video was systematically analyzed using a combination of low-level and high-level visuospatial metrics (see Section \ref{sec:profiling}). By integrating these annotations, we computed a composite complexity score for each scene, enabling us to place each one along a graded continuum from low to high complexity (Figure \ref{fig:model_scale}). This approach allows for a nuanced understanding of complexity that incorporates not only visual clutter but also high-level semantics, interactivity, and multimodal cues.

Notably, clutter and structural similarity (SSIM) metrics are also inversely proportional, as one increase when the other decreases. Scenes with high clutter typically exhibit lower SSIM values. However, our results also highlight that comparable levels of overall complexity can arise from different combinations of attributes. For example, comparing scenes 23 (Hong Kong) and 25 (Bombay), we observe that while scene 23 has higher clutter values, scene 25 involves more motion, human interaction, and audio cues—factors that likely contribute to a higher perceived complexity in scene 25. This suggests that visuospatial complexity is not determined by any single attribute but by the interaction of multiple visual and semantic elements. The analysis underscores the importance of a multidimensional framework for characterizing visuospatial complexity. While clutter can impact attentional allocation, dynamic features such as motion and social interactions engage predictive, interpretive, and response-related cognitive processes. A richer taxonomy of complexity, grounded in both low- and high-level features, enables more precise evaluation of AI perception systems and supports the design of human-centered environments. To further examine how these complexity attributes affect behavior, we conducted a controlled VR driving experiment (see Section \ref{sec:study}). This allowed us to assess how different complexity profiles influence driver attention, decision-making, and performance in a fully embodied, multimodal setting.

\subsection{The effect of Visuospatial Complexity in Active Vision: An empirical VR study} \label{sec:study}

A deeper understanding of how humans perceive and act in dynamic, real-world environments requires behavioral studies that go beyond static 2D scene experiments. Real-world scenes involve multiple interacting dimensions of visuospatial complexity—such as motion, object density, auditory cues, and social interactions—that are difficult to isolate and test systematically in traditional laboratory settings. These attributes often co-occur and interact in ways that pose analytical challenges, making it difficult to identify statistically significant effects of individual variables or their combinations on perception and behavior. To address this, we employed immersive VR, which preserves embodied experience and active vision while enabling precise experimental control. In this context, we used the visuospatial complexity model as a generative framework to create controlled, interactive driving scenarios that varied systematically in complexity across multiple dimensions—including clutter, object size, motion, and human-human interactions. These VR scenes were used to test human behavior and psychophysiological responses, including eye-tracking metrics. In an empirical study, we implemented the model in a naturalistic, interactive VR study using a driving simulator. We investigate whether a structured, multimodal characterization of complexity could predict perceptual performance and attentional dynamics in real-time behavior. The model allowed us to design and parametrize the experimental environment using a range of complexity attributes: quantitative (e.g., object density), structural (scene layout, symmetry), dynamic (motion, flicker), and auditory (noise, audio events). Based on combinations of these features, we constructed levels of overall visuospatial complexity: low, medium, and high (Fig. \ref{fig:driving}).

\begin{figure}[t]
\centering
\includegraphics[width=1\textwidth]{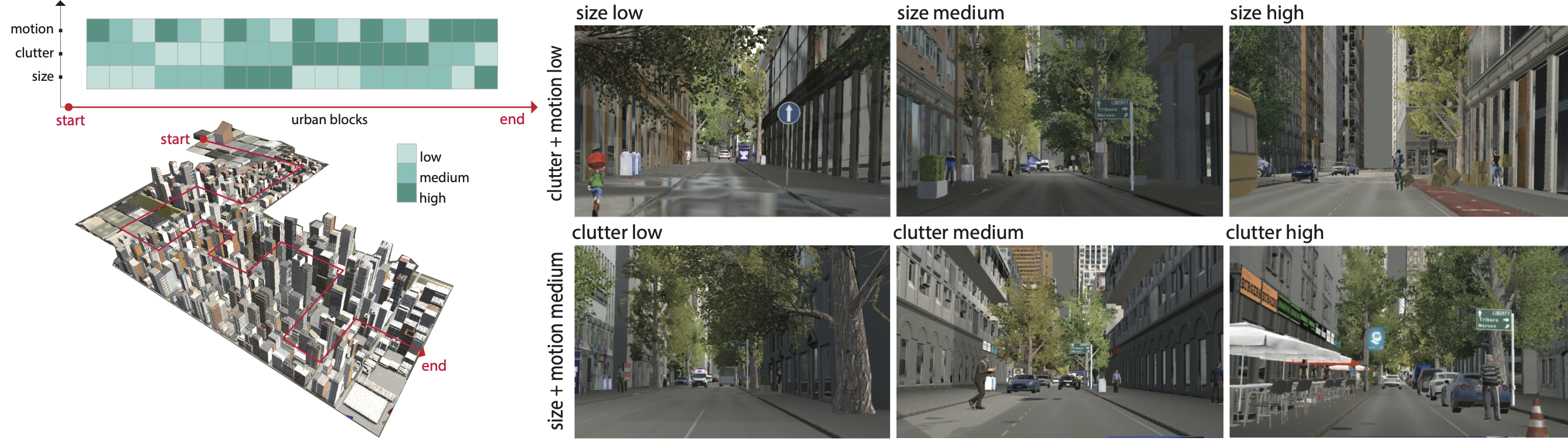}
\caption{The setup and the virtual environment of the virtual driving study. The environment involves different levels of visuospatial complexity based on the combination of visuospatial complexity attributes presented in Table \ref{tbl:visual complexity model}.} 	  
\label{fig:driving}
\end{figure}

\smallskip

\textbf{Visuospatial Complexity and Change Detection.} \quad Participants engaged in a change detection task while they navigate a virtual city through a VR simulator.  They were asked to detect changes in the behaviour of virtual agents or in other objects in the scene. Some changes were relevant to the driving task (e.g. zebra crossing, hazard situation) and other were not. The drivers had to engage in some of these events by changing velocity, deviating from their route, keeping situation awareness and alertness. We also manipulated temporal complexity by varying the interval between successive events, creating six levels (0–8 seconds) to assess the effects of temporal pressure and attentional load. A total of 80 participants (59 male, 21 female; ages 17–45) navigated through a virtual city. Results showed that visuospatial complexity significantly influenced participants’ gaze behaviour, change detection accuracy and reaction times (see \cite{Kondyli2023CRPI}). Specifically, scenes with high visuospatial complexity in terms of quantitative, structural, dynamic cues led to reduced detection performance and delayed responses. However, when participants were engaged with highly engaging interactions, task-relevant events, or hazardous situations, the detrimental effects of complexity were mitigated. This interaction effect highlights an important finding: the perceptual cost of  complexity depends on the combination of attributes at present, is not fixed but is modulated by the nature of the user’s embodied, goal-directed engagement with the surrounding environment. Without the structured guidance provided by the visuospatial complexity model, isolating and analyzing such interactions in a controlled yet ecologically valid setting would not have been feasible.
\smallskip

\textbf{Temporal Complexity and Change Detection.} \quad A second key finding emerged from the manipulation of temporal complexity. Detection performance was significantly impaired when two events occurred simultaneously (0-second interval), with only a 24.8\% detection rate and an average reaction time of 1.633 seconds  \cite{Kondyli2024SAP}. Performance improved substantially for event durations of 1–5 seconds but plateaued thereafter. This suggests a non-linear sensitivity threshold: attentional readiness is particularly vulnerable to immediate temporal overload, but stabilizes once a minimal processing interval is provided. In summary, this study demonstrates the practical utility of the visuospatial complexity model not only as a theoretical framework but also as a tool for designing and analyzing behavioral experiments in dynamic, embodied contexts. The model enables structured manipulation of complex environmental variables, offering a foundation for future research in human-centered AI, adaptive driver assistance systems, and cognitive modeling in real-world tasks.

\begin{figure}[t]
\centering
\includegraphics[width=1\textwidth]{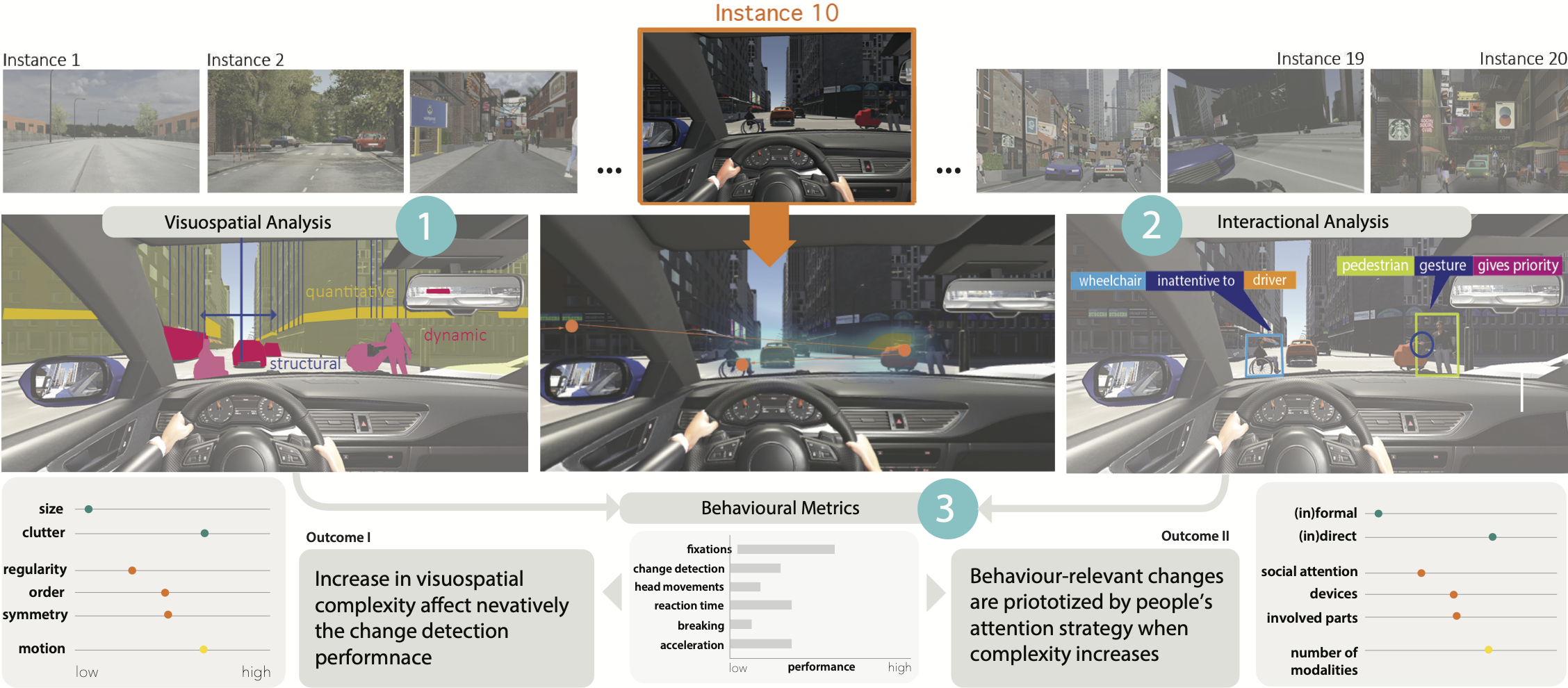}
\caption{Example of benchmarking human-factors driving dataset based on the cognitive principles of  visuospatial complexity.}
\label{fig:benchmarking}
\end{figure}

\section{Outlook}\label{sec:outlook}

This work lays the foundation for future research on assessing complexity in diverse domains --such as cognitive robotics, autonomous driving, social communication-- by refining our understanding of how complexity is constructed and perceived in real-world, multimodal environments. This understanding is essential for developing adaptive, intelligent systems that can reason, interact, and operate effectively in human-centred contexts \cite{AVI-cogvis}. Building on this foundation, we plan to pursue two lines of work:

\textbf{Qualitative human evaluation.} \quad We are conducting an empirical human-subject study to evaluate how people subjectively perceive and rank visuospatial complexity. Participants will assess a systematically parameterized set of stimuli, providing both quantitative ratings (e.g., on a scale) and qualitative distinctions. This empirical foundation will allow us to align computational measures of complexity with human perceptual experience. A qualitative analysis is essential, as visuospatial complexity is inherently subjective and culturally influenced—making it difficult to capture through quantitative metrics alone. Moreover, physiological measurements and statistical modeling gain explanatory power when anchored in participants' subjective interpretations and perceptual judgments. 

\textbf{Benchmarking for Human Factor Assistive Technologies.} \quad  Our visuospatial complexity framework offers a foundation for benchmarking in diverse machine learning (ML) and cognitive vision scenarios that involve embodied human experience and human-machine interaction \cite{DBLP:conf/kr/SuchanBM25}. Existing benchmarks often focus on dataset scale or sensor variety, but rarely account for the complexity and variability of real-world environments—factors essential for robust and generalizable AI. By systematically characterizing visuospatial complexity across datasets, our model enables the identification of underrepresented conditions and combinations of scene attributes (Fig. \ref{fig:benchmarking}). This supports more meaningful evaluation of perceptual and interaction demands in embodied contexts. The framework is particularly well-suited for domains such as autonomous driving and assistive robotics, where cognitive demands arise from rich, dynamic, and multimodal environments.

This work is conducted in synergy with ongoing research in computational cognitive vision, particularly the development of integrated vision and semantics for active, explainable visual sensemaking in autonomous vehicles \cite{DBLP:conf/kr/SuchanBM25,Suchan2021}. This empirically grounded model of complexity serves as the basis for a declarative computational framework that combines knowledge representation with visual processing to support systematic complexity analysis \cite{Kondyli2022KR}. Ultimately, our visuospatial complexity-based framework provides a scalable and cognitively grounded foundation for the design, evaluation, and benchmarking of autonomous systems. By embedding human-centred considerations—such as perceptual complexity—into dataset construction and evaluation metrics, we move toward AI systems that are not only functionally capable but also explainable, ethically aligned, and perceptually attuned to the environments in which they operate.

	        {
			\bibliographystyle{unsrt}

        	}

\end{document}